%Paper: gr-qc/9403009
%From: jemal@roxanne.nuclecu.unam.mx (Jemal Guven)
%Date: Wed, 2 Mar 1994 12:28:18 -0600

%%%%%%%%%%%%%%%%%%%%%%%%%%%%%%%%%%%%%%%%%%%%%%%%%%%%%%%%%%%%%%%%%
%%%%%%%%%%%%%%%%%%%%% Plain TeX file %%%%%%%%%%%%%%%%%%%%%%%%%%%%%%%%
%%%%%%%%%%%%%%%%%%%%%%%%%%%%%%%%%%%%%%%%%%%%%%%%%%%%%%%%%%%%%%%%%

\magnification=1200
\baselineskip=20pt
\centerline{\bf THE ISOLATION OF GRAVITATIONAL INSTANTONS:}
\centerline{\bf FLAT TORI vs. FLAT $R^4$}
\vskip1pc
\centerline {\bf Jemal Guven}
\vskip1pc
\it
\centerline {Instituto de Ciencias Nucleares }
\centerline {Universidad Nacional Aut\'onoma de M\'exico}
\centerline {A. Postal 70-543. 04510 M\'exico, D. F., MEXICO} \rm
\centerline{(guven@roxanne.nuclecu.unam.mx)}
\vskip1pc
\centerline{\bf Abstract}
\vskip1pc
{\leftskip=1.5cm\rightskip=1.5cm\smallskip\noindent
The role of topology in the perturbative solution of the Euclidean
Einstein equations (EEEs) about flat instantons is examined.
When the topology is open (with asymptotically flat boundary conditions)
it is simple to demonstrate that all vacuum perturbations
vanish at all orders in perturbation theory;
when the topology is closed (a four-torus say)
all but a 10-parameter family of global
metric deformations (moving us from one
flat torus to another) vanish.
Flat solutions, regardless of their topology, are
perturbatively isolated as solutions of the EEEs.
The perturbation theory of the complete
Einstein equations contrasts dramatically with that of the
trace of these equations, the vanishing of the scalar curvature.
In the latter case, the flat tori are
isolated whereas $R^4$ is not. This is a
consequence of a linearization instability of the
trace equation which is not a linearization instability
of the complete EEEs. \smallskip}
\vfill
\eject
\noindent{\bf I. Introduction}
\vskip1pc

When the spatial topology is closed
the solution of the (Lorentzian) Einstein equations (LEEs)
is always complicated by the requirement that
the solution be consistent with various integrability
conditions associated with the closure of the topology.

An unexpected uniquely perturbative manifestation of these integrability
conditions was discovered twenty one years ago in a pioneering
paper by Brill and Deser[1,2,3]:
when the LEEs are perturbed about any
spatially closed solution which admits a
killing vector integrability conditions
appear at second order in perturbation theory
which restrict the function space of the linear
perturbations --- the equations  suffer from a linearization
instability. The important point is that these constraints cannot
be derived by examining the linearized equations alone.
Thus, not only is the validity of the linearized theory
undermined but the implementation of perturbation theory becomes
seriously complicated. Indeed, the consequences
of the linearization instability of the Einstein
equations extend beyond perturbation theory.
For they imply the impossibility, even in principle, of
identifying the physical phase space of general
relativity when the topology is closed [4].

The linearization instability of the LEEs (the subject of Ref.[2])
implies that a flat three-torus $T^3$ (corresponding to the
spacetime topology $T^3\times R$) does not admit any maximal
first order metric perturbations that are not also flat.
This means that there are no nearby solutions of the Einstein equations
possessing a maximal surface that are not also flat --- the flat
tori are isolated with respect to such perturbations.
There is no analogous obstruction to the existence of such spacetimes
when space is open and the asymptotically flat
boundary conditions which are then appropriate are imposed. The
closed spatial topology in this case severely
limits the possible nearby solutions of the LEEs.

Intuitively, one might expect the analogs
of the integrability conditions exhibited in the
Lorenzian theory to be even stronger in the Euclidean
theory and all the more so when the four-topology is closed.
This is because the Euclidean Einstein
equations (EEEs) we are now solving are elliptic,
not hyperbolic, and we would expect additional integrability conditions
to be associated with the sewing up of the topology in
the timelike direction. Indeed, the solution space of
the vacuum EEEs with a fixed closed
topology is typically found to be finite dimensional[5].

This intuition is not, however, as sound as it might appear because
there are also preciously few asymptotic
flat solutions with simple topology. To begin with,
the unique solution with $R^4$ topology
is flat $R^4$[6]. Any other asymptotically flat instanton must
be topologically non trivial. Neither is there any
not-flat solution with topology $R^3\times S^1$[7,8].
In both cases, the key component underpinning the
proof is the positivity of the ADM mass.
There are also good reasons to believe that the only solutions with
topology $R^2\times S^2$ are the one-parameter family of
Schwarzschild instantons corresponding to a positive  ADM
mass. We must conclude that the finite
dimensionality of the solution space of the EEEs is
not a unique feature of closed topologies.
Asymptotic flatness can be an even more
restrictive boundary condition than
closure on solutions of the EEEs.

In this paper we examine the EEEs perturbatively
with a view to identifying the source of this
apparent disregard for topology. We begin in sec.II by
contrasting perturbation theory about two flat solutions of these
equations, flat $R^4$ with asymptotically flat boundary conditions
--- open in all directions, and a flat four-torus $T^4$ which is
completely closed.

The only allowed perturbations of flat $T^4$ are perturbations moving
us about the ten parameter family of globally inequivalent flat $T^4$s[9].
In fact, we can show that modulo these, perturbations
vanish to all orders in perturbation theory. There do not appear,
however, to be any obstructions on $T^4$ limiting the solution of
the full EEEs to one of these flat solutions. Such solutions
are strictly perturbatively isolated from the given flat solution.
This contrasts with the open topology where the flat solution is unique.

The four-tori are the Euclidean analogs of the flat
spacetimes with topology $T^3\times R$ in the Lorenzian theory.
One might therefore have expected some analog of Brill and Deser's
linearization instability  to show up. However,
linearization instability plays no role in establishing the isolation of
flat tori. No (non-trivial) first order pertubations survive the
integrability conditions operating at that
order in the Euclidean theory. If we had, however, failed
to recognize these conditions at first order, the
integrability conditions appearing at second order
would have mopped up the spurious solutions resulting from
our oversight.

To provide a heuristic argument why one would expect linearization
instability to play a diminished role in the Euclidean Einstein theory,
in sec.III we examine briefly what happens to the
linearization instability of the Lorenzian theory studied by
Brill and Deser when the signature of the spacetime metric is
made Euclidean. What occurs is that the quadratic form in first
order quantities which
appears at second order in perturbation theory is no longer
positive definite. The result is that the associated
integrability condition is a much weaker one.

In sec.IV we examine a subset of the vacuum EEEs,
the vanishing of their trace (the scalar curvature). Now,
every solution of the full theory is clearly also a solution of this
truncated theory but not conversely. However, unlike the full EEEs, this
single equation does suffer from a linearization instability which
kills all non flat TT perturbations of any flat $T^4$. One
{\it does} have to proceed to second order in perturbation to see this.
The analogs of these perturbations satisfying asymptotically flat
boundary conditions not only survive on $R^4$ but
are freely specifiable. The essential
difference between the EEEs and the trace equation in
this context is that solutions of the latter equation do
not necessarily  satisfy the positive mass (action) theorem which is
a property of the full Einstein equations and not of any proper
subset thereof.

This result has a simple application.
If linearization instability kills a perturbation in the
truncated theory it clearly also kills it in the full theory
whether or not the full theory itself suffers from a
linearization instability. We exploit this reasoning to
prove the perturbative isolation of the EEEs
coupled to matter with a trace negative stress tensor.

\vskip2pc
\noindent{\bf II. Euclidean Einstein Equations: $R^4$ vs. $T^4$}
\vskip1pc

Let us begin by examining the ingredients contributing to the
perturbative isolation of flat instantons
in greater detail. We decompose the metric about the flat background

$$g_{ab}=\delta_{ab}+\epsilon h_{ab}+
\epsilon^2 j_{ab}+{\cal O}(\epsilon^3)\,.\eqno(1)$$
The symmetric tensor $h_{ab}$ ($j_{ab}, \cdots$)
can be decomposed as follows

$$\eqalign{h_{ab}=& h^{t}_{ab}+
\partial_a h^L_b+\partial_b h^L_a\cr
=&h^{TT}_{ab}+ {1\over3}(\delta_{ab}\Delta-\partial_a\partial_b)h^T
+{1\over4}\delta_{ab}C_h + \partial_a h^L_b+\partial_b h^L_a\,,\cr}\eqno(2)$$
where $\Delta\equiv\partial^a\partial_a$.
The first line represents the decomposition into a transverse
component $h^t_{ab}$ and a longitudinal component characterized by a
vector, $h^L_a$.
The divergence of $h_{ab}$ determines $h^L_a$
up to an irrelevant killing vector of the background geometry.
On the second line, the transverse component is itself decomposed
into a transverse traceless component, $h^{TT}_{ab}$ and a remainder
characterized by the scalar $h^T$ and the constant $C_h$, which is transverse
by construction. The constant mode (which only appears
when the topology is closed) is the spatial average of $h^{ta}{}_a$,

$$C_h={1\over V}\int dV\,h^{t\,a}{}_a\,.\eqno(3)$$
At first order the Ricci tensor is given by

$$R^{(1)}_{ab}={1\over 2}\left(-\partial_a\partial_b h-\Delta h_{ab}+
\partial^c\partial_b h_{ca}+\partial^c\partial_a h_{cb}\right)\,.
\eqno(4)$$
The linearized trace equation

$$R^{(1)}=-\Delta h
+\partial^a\partial^b h_{ab} =0
\eqno(5)$$
is identically satisfied by $h^{TT}_{ab}$, $h^L_a$ and $C_h$. It reduces
therefore to an equation determining $h^T$:
$$\Delta^2 h^T=0\,.\eqno(6)$$
The only solution of Eq.(6) consistent with the open topology is
$h^T=0$; with the closed topology, $h^T={\rm constant}$. This constant
can, however, safely be dropped as it does not contribute to
$h_{ab}$. It is a special feature of perturbations about flat space that
the equation $R^{(1)}=0$ can be solved without
reference to any gauge conditions. More generally, the
equation will not be tractable until gauge conditions freezing the
diffeomorphism invariance of the theory are
specified and the resulting equations solved simultaneously.
The appropriate variables to gauge fix are $h^L_a$.
We choose the four gauge conditions

$$h^L_a=0\,.\eqno(7)$$
In perturbation theory, it is obvious that
the `natural' gauge choice is just the gauge, (7).

After gauge fixing, we are left with the unconstrained variables
$h^{TT}_{ab}$ and $C_h$.

The perturbed tracefree vacuum EEEs are now given by

$$\Delta h^{TT}_{ab}=0\,,\eqno(8)$$
which is Laplace's equation for each $h^{TT}_{ab}$.
We should have anticipated that the linearized
Euclidean equations would assume the form (8):
The Laplace equation is nothing other than the analytic continuation of the
wave equation we obtain when we linearize the
Lorenzian Einstein equations. The only everywhere regular solution
consistent with asymptotically flat boundary conditions
$h^{TT}_{ab} \sim {\cal O}(r^{-1})$ on $R^4$
is $h^{TT}_{ab}=0$; in the case of $T^4$ the
appropriate boundary conditions are periodic conditions
on $R^4$ treated as the universal cover of $T^4$. The solution is
a linear combination of the $9$ constant TT-perturbations. With $C_h$,
these make up a ten parameter family of metric
perturbations of $T^4$. The Riemann curvature vanishes on any
of these perturbations. These solutions are
the perturbative relics of the
ten dimensional space of parameters characterizing the global geometry
of flat $T^4$'s. These parameters can be taken to be the
lengths of the four homotopically inequivalent closed minimal
curves on $T^4$ and the six angles between them.
We therefore can conclude that the only allowed infinitesimal
physical (non-diffeomorphism) metrical
deformations of $T^4$ are changes in this parameter space.
These deformations survive because the
EEEs do not distinguish between globally inequivalent
flat spaces with the same topology. We note that

$$C_h=2{\delta V\over V}\,,$$
when $h^T=0$ and $h^L_a=0$ and therefore
corresponds to a change of volume.
This can be constructed from an appropriate linear combination of the
ten parameters.

At second order,
note that the contribution to $R^{(2)}_{ab}$ due to $h_{ab}$ is given by [10]

$$\eqalign{R^{(2)}_{ab}=&{1\over 2}\bigg[
{1\over2}\partial_a h^{cd}\partial_b h_{cd}+h^{cd}
(\partial_a\partial_b h_{cd}+\partial_c\partial_d h_{ab}-
\partial_b\partial_d h_{ac}-\partial_a\partial_d h_{bc})\cr
&+\partial^d h_b{}^c(\partial_d h_{ac}-\partial_c h_{ad})-
(\partial^d h^c{}_d-{1\over 2}\partial^c h)(\partial_b h_{ac}+\partial_a
h_{bc}-
\partial_c h_{ab})\bigg]\,.\cr}\eqno(9)$$
This term acts as a source for
the second order perturbation, $j_{ab}$.
However, $h_{ab}$ is zero (constant)
so that $j_{ab}$ and all higher orders vanish (are also constants).

\vskip2pc
\noindent{\bf III Linearization Instability: LEE's vs. EEEs}
\vskip1pc
The isolation of the flat tori as solution of the
the EEEs is not a consequence of a
linearization instability in these equations. Indeed, if there is
any linearization instability in the EEEs associated with the
closure of the topology there are good reasons for
believing that it is not as severe as the
linearization instability in the corresponding
Lorenzian theory. As a justification for this claim
consider a Euclidean retreatment of Brill and Deser's perturbative
ADM analysis of the three-dimensional flat torus with
spacetime topology $T^3\times R$. In the corresponding
Euclidean problem with the same topology,
the change in the spacetime signature manifests itself in a change of
sign of the extrinsic curvature quadratic in the
Hamiltonian constraint[10]. Eq.(5.4) of Ref.[2] gets replaced by (also
see below)

$$\Delta^2 j^T- (\pi^{TT\,ab})^2+
{1\over4}(\vec\partial h^{TT}_{ab})^2
+{\rm divergence}=0\,.\eqno(10)$$
The linearization instability (if there is one)
is realized by integrating this
equation over the closed
spatial volume. The divergence terms involving both
first order and second order quantities now vanish.
We get

$$\int dV \left[(\pi^{TT\,ab})^2 -
{1\over4}(\vec\partial h^{TT}_{ab})^2\right]=0\,.\eqno(11)$$
The Euclidean constraint Eq.(11)
is considerably weaker than its Lorenzian counterpart (where the
sign of $(\vec\partial h^{TT}_{ab})^2$ is reversed). In the
latter case the positive definiteness of
the quadratic in $\pi^{TT\,ab}$ and $h^{TT}_{ab}$
implied the extremely stringent condition
$\pi^{TT\,ab}=0=\partial_c h^{TT}_{ab}$ on the first order
perturbations.\footnote {$^1$} {We are ignoring the
fact here that if we consider the full set of linearized EEEs,
both $\pi^{TT\,ab}$ and $h^{TT}_{ab}$ vanish at first order. Therefore
Eq.(11) is not a manifestation of a linearization instability.}

\vskip2pc
\noindent{\bf IV Perturbations of Zero Ricci Curvature vs.
Perturbations of Zero Scalar Curvature}
\vskip1pc
The way we solved the perturbative vacuum EEEs was
to first solve the non-linear elliptic equation,

$$R=0\,,\eqno(12)$$
perturbatively.	Flat $R^4$ admitted non trivial perturbative solutions,
any $h^{TT}_{ab}$ consistent with asymptotically flat boundary conditions
was a solution (it was the remaining nine tracefree equations
which annihilated them as solutions of the EEEs).
What is surprising therefore in light of this
is  that the flat $T^4$'s do not admit
non-flat perturbative solutions of Eq.(12).  This  is a
consequence of a genuine linearization instability
of the single equation, Eq.(12).

To see this, we note that Eq.(12) on
any space with $T^4$ topology
is exactly the super-Hamiltonian constraint in five dimensional
general relativity on a momentarily static slice through
a spacetime with topology, $T^4\times R$.
This means that the problem has been conveniently
reduced to a special case of Brill and Deser's
analysis (stepped up one dimension)
and we can exploit the techniques used in Refs.[1] and [2].
For completeness, however, and for the purpose of
making various observations we will reproduce some of the details.

We note that in two dimensions, the only solutions to Eq.(12)
are flat. In all higher dimensions,
however, this equation admits non-flat solutions.
The existence of non-trivial solutions to this
equation assures us that the perturbative isolation we discuss
is not merely a rediscovery  of a non-perturbative isolation
which is rendering the closed flat solution unique in any case.

To examine Eq.(12) perturbatively we expand
the metric about the flat background $\delta_{ab}$ in the manner of Eq.(1).
The game is to identify integrability conditions which appear
at each order in perturbation theory which kill perturbations
of the preceeding order. For the purpose of simplifying
the statement of these perturbative integrability conditions,
it is more convenient to consider perturbations of the density,
$\sqrt{g}R$. Let us first recall various elementary properties of the
density, $\sqrt{g}R$. Corresponding to any first order local variation
of the metric, $g_{ab}\to g_{ab}+\delta g_{ab}$

$$\delta\left[\sqrt{g}R\right]=-\sqrt{g}G^{ab}\delta g_{ab}
+\sqrt{g}g^{ab}\delta R_{ab}\,.\eqno(13)$$
The important point is that the second term is a divergence.
This feature will be extremely useful in that it enables us to
isolate divergences in the perturbative
expansion of $\sqrt{g}R$.
In the absence of boundary surfaces, these divergences will be killed
when we integrate over the manifold.

Let us denote the $n^{th}$ order
perturbation in $\sqrt{g}R$ by $\left(\sqrt{g}R\right)^{(n)}$.
In particular, at the first order in perturbation
theory about flat space the first term on the RHS of Eq.(13) vanishes and
$\left(\sqrt{g}R\right)^{(1)}$ is the divergence appearing in Eq.(5).
As before, the solution to $\left(\sqrt{g}R\right)^{(1)}=0$ is
(without any loss of generality) $h^T=0$.
We fix the gauge as before with Eq.(7) and we are left with the
unconstrained variables $h^{TT}_{ab}$ and $C_h$.

When we proceed to the
next order of perturbation theory, an integrability condition will
appear which completely kills $h^{TT}_{ab}$ when the space is closed.
Let us examine the second order equation
$\left(\sqrt{g}R\right)^{(2)}=0$ a little more carefully to see how
this comes about.
To simplify the evaluation of $(\sqrt{g}R)^{(2)}$, we
express the second order variation in the form

$$\delta^2\left(\sqrt{g}R\right)=
-\delta\left(\sqrt{g}G^{ab}\delta g_{ab}\right)
+\delta(\sqrt{g}g^{ab}\delta R_{ab})\,.\eqno(14)$$
What makes things simple is that the divergence term will remain a divergence
under the second variation. Thus
$\delta(\sqrt{g}g^{ab}\delta R_{ab})$ is another divergence.
Therefore, at second order

$$\left(\sqrt{g}R\right)^{(2)}
=-\left(h^{ab}-{1\over2}h\delta^{ab} \right)R^{(1)}_{ab}
+{\rm divergence}\,,
\eqno(15)$$
where the first order perturbation in the Ricci tensor is given by
Eq.(4). We note that, modulo the divergence, the second order
perturbation in the density is independent of the second order
perturbation in the Ricci tensor, $R^{(2)}_{ab}$.\footnote *
{This device was exploited in Refs.[1] and [2] although its
potential was never emphasized there.}

We can continue in this way to conclude that
$\left(\sqrt{g}R\right)^{(n)}$ involves only the preceeding order
perturbation in the Ricci tensor. Consultation of Eq.(9) should be
sufficient to appreciate the value of grouping terms this way. We have not
explicitly written down the contributions due to the second order
perturbations, $j_{ab}$  which appear at second order.
It is however a	 simple matter to incorporate them at this level.
The contribution to the density at second order
will be the corresponding first order expression with $h_{ab}$
replaced by $j_{ab}$. Thus the second order contributions due to $j_{ab}$
is the divergence appearing in Eq.(5) with $h_{ab}$ replaced by $j_{ab}$.
This is an important yet scarcely explored feature of perturbative
canonical Einstein gravity (in a technical sense, it is the origin
of the linearization instability constraint).
The second order constraint equation is

$$\Delta^2 j^T+
{1\over4}(\vec\partial h^{TT}_{ab})^2
+{\rm divergence}=0\,,\eqno(16)$$
where the divergence appearing in Eq.(16) involves only  quadratics
in first order perturbations.
Only the $j^T$ component of the second order metric perturbation, $j_{ab}$
appears.
Eq.(16) is therefore an equation for $j^T$ with source consisting of
terms quadratic in first order quantities.
The remaining components of $j_{ab}$
remain undetermined at second order.
$C_h$ does not appear in the equation.
Because the space is closed, the integral of Eqs.(16)  over
the spatial volume yields the constraint (linearization
instability) on the first order
perturbations:

$$\int dV (\vec\partial h^{TT}_{ab})^2
=0\,.\eqno(17)$$
The first order transverse traceless perturbations must therefore, as
before, be constants.
The only perturbations which survive are the
perturbations corresponding to the ten parameters
discussed above.

The two essential features contributing to the result we have
established are that the background space is both closed and flat.
There is no corresponding result when space is open and flat.
If the background is not flat, $j_{ab}$ will
appear in terms other than the divergence. These terms survive
the integration over the volume of space, and the equation
corresponding to Eq.(17) no longer represents a constraint on the first
order perturbations.

At higher orders in perturbation theory the pattern will be repeated
with constraints appearing on the higher order perturbations killing
all but the ten constant deformations discussed earlier.
It is therefore impossible as before to perturb away from a
flat solution. This differs from the
corresponding result for the full EEEs is that the
obstruction here can be attributed to topology.

\vskip2pc
\centerline{\bf V. An Application}
\vskip1pc

If a linearization instability kills a perturbation in the
truncated theory it clearly also kills it in the full theory
whether or not the full theory itself suffers from a
linearization instability. This reasoning can be
exploited to prove that the flat tori are
isolated as solutions of the EEEs coupled to matter with a trace-free
stress tensor (in particular, the vacuum theory considered in sec.II)

$$T^a{}_a = 0\,.\eqno(18)$$
Such a stress tensor corresponds to a conformally invariant material
field. The trick is to note that whenever Eq.(18) holds
any solution of

$$G_{ab}=8\pi T_{ab}\eqno(19)$$
also satisfies Eq.(12).	It is also clear using an
identical argument to that presented in Ref.[1]
with the energy density replaced by $T^a{}_a$ that
this result also goes through under the weaker
condition that $T^a{}_a$ ($R$) be negative (positive) definite.

We stress that it is incorrect to conclude from this that the flat tori
are isolated in this way because of a linearization instability
of the complete  EEEs. The linearization instability
we have exploited is that of the single traced equation.
The linearization instability in this
subset of the field equations does, however, eliminate the
necessity to consider the remaining equations. The trade off is that
one needs to proceed to second order in perturbation theory.
Whether this is a peculiarity of perturbation theory
about flat solutions or can be exploited in other contexts
remains to be seen.

One final comment on the example considered above. Does the
addition of the source of matter
considered above in any way change our conclusions for
the vacuum theory? To test if it does one needs to know
how flat $R^4$ respond to perturbations
coupled to a stress tensor of the form considered above.
The answer is no. For the non-existence result
in the vacuum theory remains valid
when the Ricci tensor is positive
(in particular, if $T_{ab}$ has a positive trace)[11].

With a generic source of matter, solutions of the
asymptotically flat theory will exist. On the closed
topology, however, they will be subject to
integrability conditions.

\vskip2pc
\centerline{\bf VI. Conclusions}
\vskip1pc

In this paper, we have examined the dependence on topology of
perturbative solutions of the EEEs about flat instantons.
This dependence contrasts dramatically with that of
analogous Lorenzian solutions. Closure, in the
Euclidean theory is not necessarily a more
restrictive condition than asymptotic flatness.
In particular, linearization instability does not appear to play the
significant role it does in the Lorenzian theory.

The nature of the restrictions associated with the topology appears also
to depend crucially on the elliptic system of equations we
are solving. In particular, the solution space of the EEEs
and that of the trace of these equations
have very different dependences on the topology.

It is not clear if physics is sensitive to these
topological considerations. In the semi-classical context,
solutions of the Euclidean equations of motions
appear as saddle points in the functional integral
formulation of the quantum theory. The prototype is the
decay of the false vacuum calculation in a
self-interacting scalar quantum field theory with
a true and a false vacuum[12]. In the semi-classical
approximation, the tunneling amplitude from
false to true vacuum is dominated by an $O(4)$ invariant bounce.
However, there are nearby instantons which are not $O(4)$ invariant [13].
The non-existence of nearby instantons in the Einstein theory might appear to
suggest that the semi-classical approximation to quantum gravity represents
a poor approximation to the exact theory. However, the non-existence
of nearby solutions can only reduce the prefactor of the exponential
and therefore cannot affect the conclusions of the
approximation. There is another way in which
the approximation might fail, however. For even if an instanton can be
found which is consistent with some given set of boundary conditions,
there is no guarantee that a nearby instanton exists when the
boundary conditions are subjected to a small perturbation.
In particular, a tunneling channel provided by a given instanton
appropriate to a given set of boundary conditions might disappear when these
boundary conditions are perturbed.

\vskip1pc
\centerline{\bf Acknowledgements}
\vskip1pc
I would like to thank Riccardo Capovilla and
Niall \'O Murchadha for helpful comments on this work.
I have especially benefitted from correspondence with
the referee who patiently pointed out various flaws in
earlier drafts of this paper.

\vskip2pc

\noindent{\bf REFERENCES}
\vskip1pc

\item{1.} D. Brill in {\it Gravitation: Problems and Prospects}
Naukova Dumka Kiev (1972)
\vskip1pc
\item{2.} D. Brill and S. Deser  {\it Comm Math Phys} {\bf 32} (1973) 291
\vskip1pc
\item{3.} A valuable introduction to the subject (containing extensive
references) is D. Brill
{\it Linearization Instability} in {\it Spacetime and Geometry}
(ed. by R. Matzner and L. Shepley  Univ. of Texas Press, Austin, 1982)
For a summary of recent progress, see V. Moncrief in {\it Directions in
Relativity: Papers in honor of D. Brill}
ed. by T. Jacobson and B.L. Hu (Cambridge University Press, 1993)
\vskip1pc
\item{4.} C. Torre {\it Phys. Rev.} {\bf D46} (1992) 3231
\vskip1pc
\item{5.} For a summary of known results written for
mathematicians, see
A. Besse {\it Einstein Manifolds}, (Springer Verlag, Berlin, 1986)

\vskip1pc
\item{6.} E. Witten {\it Comm Math Phys} {\bf 80} (1981) 381
\vskip1pc
\item{7.} The conjecture appears in
D. Gross, M. Perry and L. Yaffe {\it Phys Rev} {\bf D25} (1982) 330
\vskip1pc
\item{8.} N. O'Murchadha and H. Shanaghan {\it Phys Rev Lett}
{\bf 70} (1993) 1576
\vskip1pc
\item{9.} J. Wolf {\it Manifolds of Constant Curvature},
(Publish or Perish, Boston, 1972)
\vskip1pc
\item{10.} C. Misner, K. Thorne and J.A. Wheeler
{\it Gravitation} Freeman, San Francisco (1973)
\vskip1pc
\item{11.} N. Bacon {\it M.Sc Thesis} (University College Cork, Ireland,
1992); For another (more arduous) proof, see
G. Jungman and R. Wald {\it Phys. Rev.} {\bf D40} (1990) 2619
\vskip1pc
\item{12.} S. Coleman {\it The Uses of Instantons} in {\it The Whys
of Subnuclear Physics}, ed. by A. Zichichi (Plenum Press, New York, 1979)
\vskip1pc
\item{13.} S. Coleman, V. Glaser and A. Martin
{\it Comm Math Phys} {\bf 58} (1978) 211

\bye